\documentclass[12pt]{iopart}
\usepackage{graphicx}
\usepackage{harvard}
\usepackage[usenames,dvipsnames]{color}

\begin{document}

\citationstyle{dcu}
\citationmode{abbr}

\title[The impact of (n,$\gamma$) reaction rate uncertainties on the i
  process]{The impact of (n,$\gamma$) reaction rate uncertainties of
  unstable isotopes near $N=50$ on the i process nucleosynthesis in He-shell flash white dwarfs}

\author{Pavel Denissenkov$^{1,2,10}$, Georgios Perdikakis$^{2,3,4}$, 
  Falk Herwig$^{1,2,10}$, Hendrik Schatz$^{2,4,5,10}$, 
  Christian Ritter$^{1,2,10}$, Marco Pignatari$^{6,7,10}$, Samuel Jones$^{8,9,10}$, Stylianos Nikas$^{2,3}$
  and Artemis Spyrou$^{2,4,5}$}
 
\address{$^1$Department of Physics \& Astronomy, University of Victoria,
       P.O.~Box 1700, STN CSC, Victoria, B.C., V8W~2Y2, Canada}
\address{$^2$Joint Institute for Nuclear Astrophysics, Center for the Evolution of the Elements, Michigan
       State University, 640 South Shaw Lane, East Lansing, MI 48824, USA}
\address{$^3$Department of Physics and Physical Sciences Program, Central Michigan University, Mt. Pleasant, Michigan 48859, USA}
\address{$^4$National Superconducting Cyclotron Laboratory, Michigan State University, East Lansing, MI 48824, USA}
\address{$^5$Department of Physics \& Astronomy, Michigan State University, East Lansing, Michigan 48824, USA}
\address{$^6$E.A. Milne Centre for Astrophysics, Department of Physics \& Mathematics, University of Hull, HU6 7RX, United Kingdom}
\address{$^7$Konkoly Observatory, Research Centre for Astronomy and Earth Sciences, Hungarian Academy of Sciences, Budapest, Konkoly Thege Mikl\'{o}s \'{u}t 15-17, 1121 Budapest, Hungary}
\address{$^8$Heidelberg Institute for Theoretical Studies, Schloss-Wolfsbrunnenweg 35, 69118 Heidelberg, Germany }
\address{$^9$Computational Physics and Methods (CCS-2) and Center for Theoretical Astrophysics, Los Alamos National Laboratory, NM 87544, USA}
\address{$^{10}$NuGrid Collaboration, \harvardurl{http://www.nugridstars.org}}
\ead{nugrid@lists.uvic.ca}

\begin{abstract}
The first peak s-process elements Rb, Sr, Y and Zr in the post-AGB
star Sakurai's object (V4334 Sagittarii) have been proposed to be the result of i-process
nucleosynthesis in a post-AGB very-late thermal pulse event.  We estimate
the nuclear physics uncertainties in the i-process model predictions to  
determine whether the remaining discrepancies with observations
are significant and point 
to potential issues with the underlying astrophysical model. We find 
that the dominant source in the nuclear physics uncertainties are predictions
of neutron capture rates on unstable neutron rich nuclei, which can have 
uncertainties of more than a factor 20 in the band of the i-process. 
We use a Monte Carlo variation of 52 neutron capture rates 
and  a 1D multi-zone post-processing model for the i-process
in Sakurai's object to determine the cumulative effect of these
uncertainties on the final elemental abundance predictions. 
We find that the nuclear physics uncertainties are large and 
comparable to observational errors. Within these uncertainties the
model predictions are consistent with observations. 
A correlation analysis of the results of our MC simulations reveals that the strongest impact on
the predicted abundances of Rb, Sr, Y and Zr is made by the uncertainties
in the (n,$\gamma$) reaction rates of $^{85}$Br, $^{86}$Br, $^{87}$Kr, $^{88}$Kr, $^{89}$Kr, $^{89}$Rb,
$^{89}$Sr, and $^{92}$Sr. This conclusion is supported by a series of 
multi-zone simulations in which we increased and decreased to their maximum and minimum limits one or
two reaction rates per run.
We also show that simple and fast
one-zone simulations should not be used instead of more realistic multi-zone
stellar simulations for nuclear sensitivity and uncertainty
studies of convective-reactive processes.
Our findings apply more generally to
any i-process site with similar neutron exposure, such as
rapidly accreting white dwarfs with near-solar metallicities.
\end{abstract}

\pacs{26.20.Fj, 25.40.Lw}

\noindent{\it Keywords\/}: nuclear reactions, nucleosynthesis, stars:
abundances, stars: AGB and post-AGB

\submitto{\jpg}

\maketitle

\section{Introduction}\label{s.introduction}

Most of the solar system chemical elements heavier than iron were
produced in the slow (s) and rapid (r) neutron capture processes in
previous generations of stars and stellar explosions
\cite{kaeppeler:11,thielemann:11}. \citeasnoun{cowan:77}
proposed that, under certain physical conditions, an intermediate (i) process with a
neutron density $N_{\rm{n}}\sim 10^{15}\, \rm{cm}^{-3}$ intermediate
between those typical for the s process ($N_{\rm{n}}\leq 10^{11}\,
\rm{cm}^{-3}$) and r process ($N_{\rm{n}}\geq 10^{20}\, \rm{cm}^{-3}$)
might be triggered in stars. Like for the main s process,
neutrons for the i process are released in the reaction
$^{13}$C($\alpha$,n)$^{16}$O. The higher neutron density in the i
process is achieved at typical He-burning temperatures $T\sim
2$\,--\,$3\times 10^8$ K due to efficient replenishment of $^{13}$C from the
reaction $^{12}$C(p,$\gamma)^{13}$N followed by the decay
$^{13}$N(e$^+\nu)^{13}$C.  This combination of He and H burning
reactions in one process occurs in a He-flash convective zone,
where there is a plenty of He and $^{12}$C, provided that a small
amount of H is ingested into it when the upper convective boundary
reaches the H-rich envelope layer. In this situation, the reactions
$^{12}$C(p,$\gamma)^{13}$N and $^{13}$C($\alpha$,n)$^{16}$O are
spatially separated, each taking place at its own favorable
conditions, and $^{13}$N with a half life of 9.96 min decays into
$^{13}$C while being carried down by convection with a comparable
turnover timescale of $\sim 15$ min.  

It was not until the late 1990s that the first strong evidence of the
i process in stars was detected.  \citeasnoun{asplund:99} investigated
the evolution of the surface elemental abundances in the post-asymptotic 
giant branch (post-AGB) star Sakurai's
object (V4334 Sagittarii) that at the time was experiencing a nova-like
transient outbreak. 
A post-AGB star is a CO core of an AGB star that has lost most of its H-rich
envelope as a result of an AGB super wind or a common-envelope event if it was in a close binary system. 
Such a star still has a He shell surrounded by a thin H envelope atop the CO core.
When the star leaves the AGB it first moves almost horizontally to the left (to the higher effective
temperatures) on the Hertzsprung-Russel diagram, as long as its luminosity is maintained by
H burning in a shell, then, when the H-shell burning dies away, the star settles down on
a white-dwarf cooling track \cite{paczynski:71}. Depending on its thermodynamic properties
and mass, the He shell can experience a very late thermal pulse (VLTP) when the star begins to cool down 
\cite{herwig:01,miller-bertolami:06}. The VLTP is a He-shell flash
similar to those occurring in thermally-pulsing AGB stars. During it, the temperature at the bottom of
the He shell rises to $\sim 2$\,--\,$3\times 10^8$ K, and the entire He shell becomes convective.
When the He-flash convection reaches the bottom of the H-rich envelope it begins to ingest
H and this triggers the i process. VLTPs are experienced by a quarter of all post-AGB stars \cite{herwig:05}.
The observed outbreak of Sakurai's object was soon interpreted as a VLTP
\cite{Herwig:2001ui}.  The observations
covered a time interval of about six months, during which the
abundances of light s-process elements, such as Rb, Sr, Y and Zr,
increased by approximately 2 dex. These abundance signatures can not
normally arise during the preceding regular AGB evolution,
and therefore provide evidence for active stellar nucleosynthesis
during the He-shell flash. According to the stellar
evolution models, a lag time of a few years (corresponding to
the thermal time scale of the remaining thin envelope layer) is expected between
the He-shell flash with its associated nucleosynthesis and the abundance
observations on the stellar surface. Therefore the observed abundance changes over the
six-month observational period probably signal mixing events rather
than ongoing nucleosynthesis.

\citeasnoun{herwig:11} suggested that both the observed changes of the
surface chemical composition of Sakurai's object and its light curve
resulted from a He-shell flash that led to the physical conditions
required for the i process, in particular, the vigorous convection
that entrains H from a thin H-rich envelope to transport it into the He burning shell. 
To reproduce the observed abundances within the
one-dimensional modeling approach adopted by \citeasnoun{herwig:11}, it
was necessary to prolong i-process nucleosynthesis by adjusting the time when the He convective zone in Sakurai's
object splits into two separate convective zones, the upper zone being
driven by burning of the entrained H. It was assumed that the split occurred at a
later time than predicted by the mixing-length theory (MLT) employed in the stellar
evolution models. Recent three-dimensional simulations of H ingestion
in Sakurai's object have revealed global horizontal oscillations of
material near the top of the He convective zone caused by the violent
burning of ingested H that resulted in a transient split of the
convective zone \cite{herwig:14}, but this is not the split that
quenches the i process, because not enough H has been burned at that
time. Further investigations of the complex three-dimensional hydrodynamic
nature of the convective-reactive He-shell flash with H ingestion in
post-AGB stars are needed to provide 1D stellar evolution and post-processing
nucleosynthesis simulations with adequate properties of the convective
He-shell split \cite{herwig:11,stancliffe:11,herwig:14}.

Recently, a number of new abundance observations have 
revealed possible i-process signatures in a broad variety of stars.
This may indicate that i-process activation
is not limited to post-AGB stars.  Among these observations are the
peculiar abundances of elements heavier than Fe in some of the carbon
enhanced metal poor stars seemingly enhanced in both, r- and s-process
elements (CEMP-r/s stars)
\cite{Beers:2005kn,masseron:10,lugaro:12,bisterzo:12}.  For some of these
CEMP-r/s stars, e.g. CS 31062-050, the observed elemental abundances
are consistent with predictions from i-process models, while the
traditional explanation of a superposition of yields from the s and r
processes shows discrepancies \cite{dardelet:14}.  The striking result
of \citeasnoun{dardelet:14} was that the observed abundances of
several CEMP-r/s stars could be reproduced using simple one-zone i-process nucleosynthesis
simulations, with constant temperature and density.  A reasonable fit
was obtained with a typical i-process neutron number density
of $N_{\rm{n}} \sim$ 10$^{15}$~cm$^{-3}$ and neutron exposure of
$\tau\sim 10$\,--\,$50$ mbarn$^{-1}$.  These results have been
recently confirmed by \citeasnoun{hampel:16}.

A possible i-process contribution has also been invoked in low-mass, low-metallicity
AGB stars, to account for the Pb deficiency observed in post-AGB stars of
low metallicity \cite{lugaro:15}, and in young open
clusters, to explain the largest [Ba/La]\footnote{We use the standard
spectroscopic notation [A/B] $ = \log_{10}(N(\mbox{A})/N(\mbox{B}))
- \log_{10}(N_\odot(\mbox{A})/N_\odot(\mbox{B}))$, where
$N(\mbox{X})$ and $N_\odot(\mbox{X})$ are abundances of an element X
in a star and the Sun.}  ratios that are neither compatible with the
s process nor with a combination of s and r processes
\cite{mishenina:15}.  Anomalous isotopic abundances possibly
originating from an i process were also found in some presolar
graphite grains \cite{jadhav:13}, as well as in presolar SiC grains of type AB
\cite{fujiya:13} and of type mainstream \cite{liu:14}.
\citeasnoun{Roederer:2016er} also proposed that the unusual [As/Ge] ratio
($+0.99 \pm 0.23$) and the enhanced [Mo/Fe] and [Ru/Fe] ratios in
the metal-poor star HD\,94028 are the signatures of an i-process
contribution with an even lower neutron exposure compared to the case
of Sakurai's object.

With such a diverse set of observational indications it is conceivable
that the i process is activated in several additional types of stars
at various evolutionary stages. Stellar evolution models have pointed to various
possible i-process nucleosynthesis sites including the He-core flash in
low-mass stars \cite{Campbell:2010eq}, the He-shell flash in a
metal-poor AGB star \cite{Fujimoto:2000fc,Herwig:2003wk,Iwamoto:2004ju,cristallo:09} or super-AGB stars
\cite{Jones:2016ex}. In addition, so-called rapidly accreting white dwarfs (RAWDs) have recently been 
proposed as possible site for an i-process \cite{denissenkov:17}. RAWDs can 
occur in close binary systems where a white dwarf accretes H-rich material 
from a main-sequence, subgiant- or red-giant-branch star at rates of the order of
$10^{-7}\,M_\odot\, \mbox{yr}^{-1}$. At these high accretion rates, H burns
stationary resulting in a growing He shell underneath a thin H envelope. 
When the He shell mass reaches $\sim 0.01-0.03\,M_\odot$, a
recurrent thermal He flash with H ingestion occurs. The nucleosynthesis is 
expected to be similar to Sakurai's object, and chemical evolution 
studies indicate that RAWDs could contribute significantly to the origin 
of Sr, Y, Zr, Nb, and Mo \cite{cote:17}.

In order to understand the role of the i process in the origin 
of the elements at various stellar sites, observed abundances
must be compared with astrophysical models for validation, to guide
model improvements, and to 
constrain the underlying physical processes and conditions. 
This approach is limited by the observational errors and the 
uncertainties in the nuclear physics that link the stellar conditions 
to a particular abundance signature. Observational 
uncertainties are known and readily available 
as part of the published data. The goal of this paper is to also 
quantify the nuclear uncertainties, to enable a meaningful 
comparison of model predictions and observations that 
reveals information about the astrophysical i-process models. 
This is important, as current  i-process models approximate complex 
mixing processes and make strong simplifying assumptions 
that need to be validated. For example, the abundance predictions from the post-AGB star model for Sakurai's object of 
\citeasnoun{herwig:11} still exhibit discrepancies with observations,
especially for Sr and Zr, but without quantifying the nuclear
physics uncertainties it is impossible to conclude whether this 
indicates issues in the underlying astrophysical model or 
the assumed stellar parameters, such as the object's mass
\cite{MillerBertolami:2007br}. 

\citeasnoun{bertolli:13} investigated the impact of nuclear uncertainties
on predicted Ba, La, and Eu abundances using a simple one-zone
i process model and found significant uncertainties of the order 
of up to 1 dex. However, they did not identify the most important nuclear reactions, but 
instead carried out their analysis in terms of the underlying nuclear theory 
assumptions used in predicting reaction rates.  
In this paper we focus on the nuclear physics uncertainties in 
predicting the Rb, Sr, Y and Zr abundances in a full multi-zone post-AGB star 
model of relevance for Sakurai's object and RAWDs, and identify the key reaction uncertainties that need to 
be improved in the future. The paper is organized as follows. 
In section \ref{init_ab}, we describe the methods and models used. 
In section \ref{physics}, we discuss the nuclear physics uncertainties and the 
resulting reaction rate variation factors.  
In section \ref{results} we present results, in particular an overall
quantification of current nuclear physics uncertainties in the predicted 
elemental abundances, and an identification of the most significant sources
of these uncertainties. The goal is to provide uncertainty information
and to guide future nuclear physics work aiming at improving these
uncertainties. A summary and conclusions are given in section~\ref{s.conclusions}.

\section{Methods and models used in our nucleosynthesis simulations}
\label{init_ab}

To characterize the nuclear physics uncertainties in post-AGB star white dwarf He 
flashes we use a multi-zone post-processing model that reproduces the 
observed abundances in Sakurai's object reasonably well. The impact 
of nuclear reaction rate variations on the predicted abundances is 
determined in a Monte Carlo framework, where reaction rates are 
varied randomly within their uncertainties. The resulting abundance 
variations define the nuclear uncertainties of the abundance predictions, 
and correlations with the nuclear reaction rate variations allow us to 
identify the most critical nuclear reaction uncertainties. To guide future work, we also explore
the appropriateness of simpler, less computationally demanding approaches
such as the use of single-zone post-processing and individual, one-by-one
reaction rate variations. The one-zone model is also used to narrow down 
the number of reaction rates varied in the Monte Carlo analysis. 

\subsection{Multi-zone simulations}

\subsubsection{Stellar post-processing model}

For the multi-zone simulations of the i process in Sakurai's object,
we use the NuGrid
multi-zone post-processing nucleosynthesis parallel code mppnp \cite{Pignatari:2016er} 
customized for the H-ingestion flash problem \cite{herwig:11}. 
The simulations take temperature $T$, density $\rho$, radius $r$ and
diffusion coefficient $D_\mathrm{conv}$ profiles from a stellar evolution model for the He convective zone in the mass
range $0.5756\leq M_r/M_\odot\leq 0.5987$ at the time of its first contact with the H-rich envelope.

\begin{figure}
  \centering
  \includegraphics[width=10cm,viewport = 5 10 450 420]{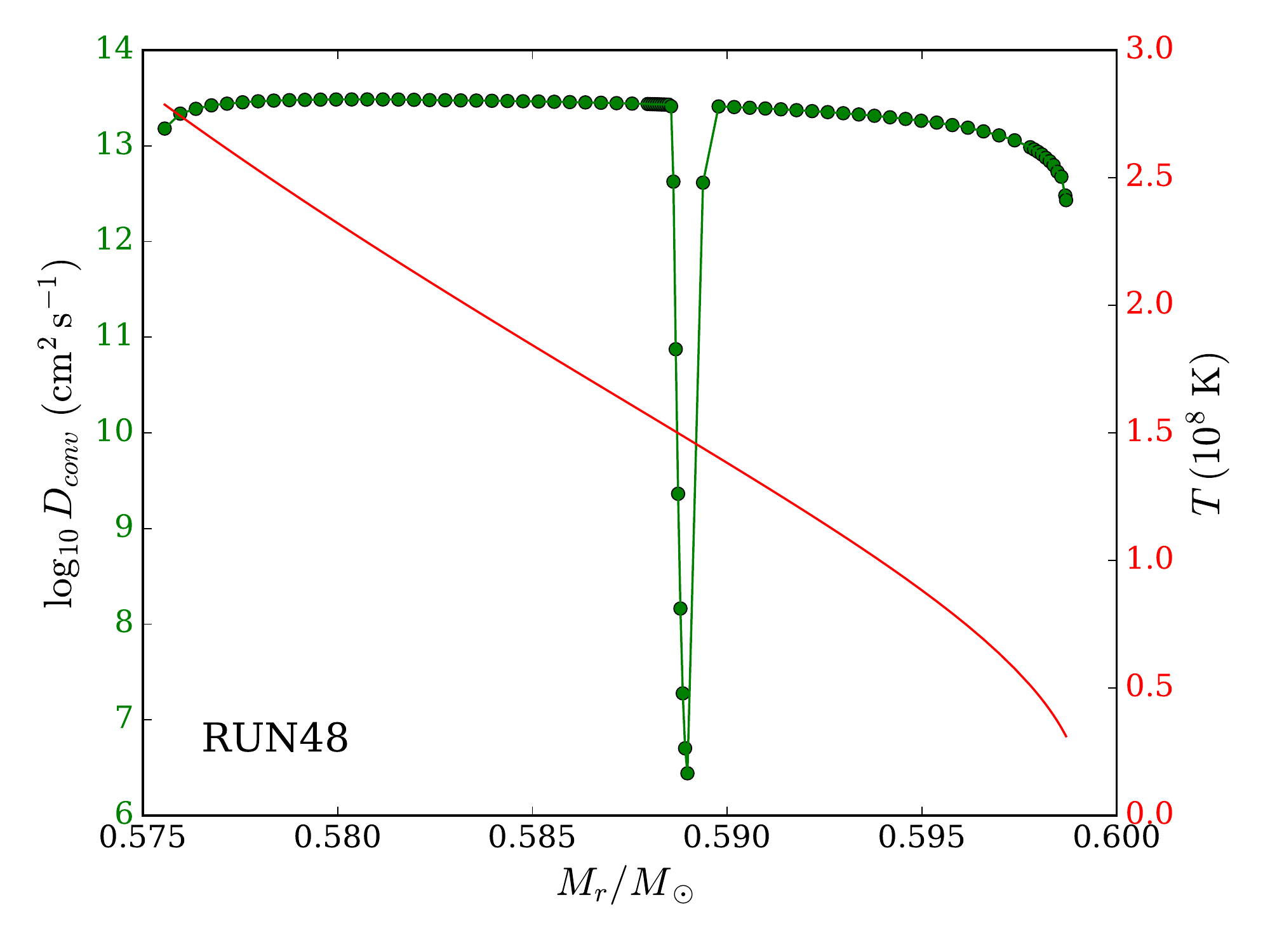}
  \caption{The MLT diffusion coefficient (green) and temperature (red) profiles in RUN48 model of
           the He-flash convective zone in Sakurai's object from \citeasnoun{herwig:11}. 
           The $D_\mathrm{conv}$ profile has a split that is artificially inserted at a specified time 
           (the 950th minute after the beginning of H ingestion) to choke off the i process by preventing
           $^{13}$C formed in the reactions $^{12}$C(p,$\gamma)^{13}$N(e$^+\nu)^{13}$C near the top of the He convective zone, where H
           is ingested, from reaching its bottom, where neutrons are released in the reaction $^{13}$C($\alpha$,n)$^{16}$O. 
           The model is divided into 81 zones (green circles), and H is ingested into the outer 5 zones.
  }
  \label{fig:fig1}
\end{figure}

In the present work, we use the same stellar evolution model ET14 of an asymptotic giant branch star of
\citeasnoun{herwig:06} during its last thermal pulse that was used by \citeasnoun{herwig:11} in their RUN48 post-processing nucleosynthesis
simulations (hereafter, RUN48 model). Its He-shell flash properties are
similar to those of Sakurai's object. \citeasnoun{herwig:11} interpolated
$T$, $\rho$, $r$ and $D_\mathrm{conv}$ for a smaller (81) than in the original ET14 model (218) number of mass zones
in the He-flash convective zone to reduce the RUN48 computation time (figure~\ref{fig:fig1}).
To reproduce the high ratio of the first to the second peak s-process elements in Sakurai's object, they
assumed that the i process in the He convective zone was choked off by its split at the 950th minute
after the beginning of H ingestion. This empirically adjusted parameter actually sets up a time period during which
the i process at the bottom of the He convective zone can change the abundances above the mass coordinate of the split.
After the split is inserted, the convective mixing above it will homogenize the abundance distributions there. 
Therefore, the final abundances at the top of the He convective zone, that are compared with the surface
chemical composition of Sakurai's object, can simply be computed by simulating the i process in this zone for 950 minutes
and then mass-averaging the resulting abundance profiles above the split.

\begin{figure}
  \centering
  \includegraphics[width=10cm,viewport = 5 10 450 420]{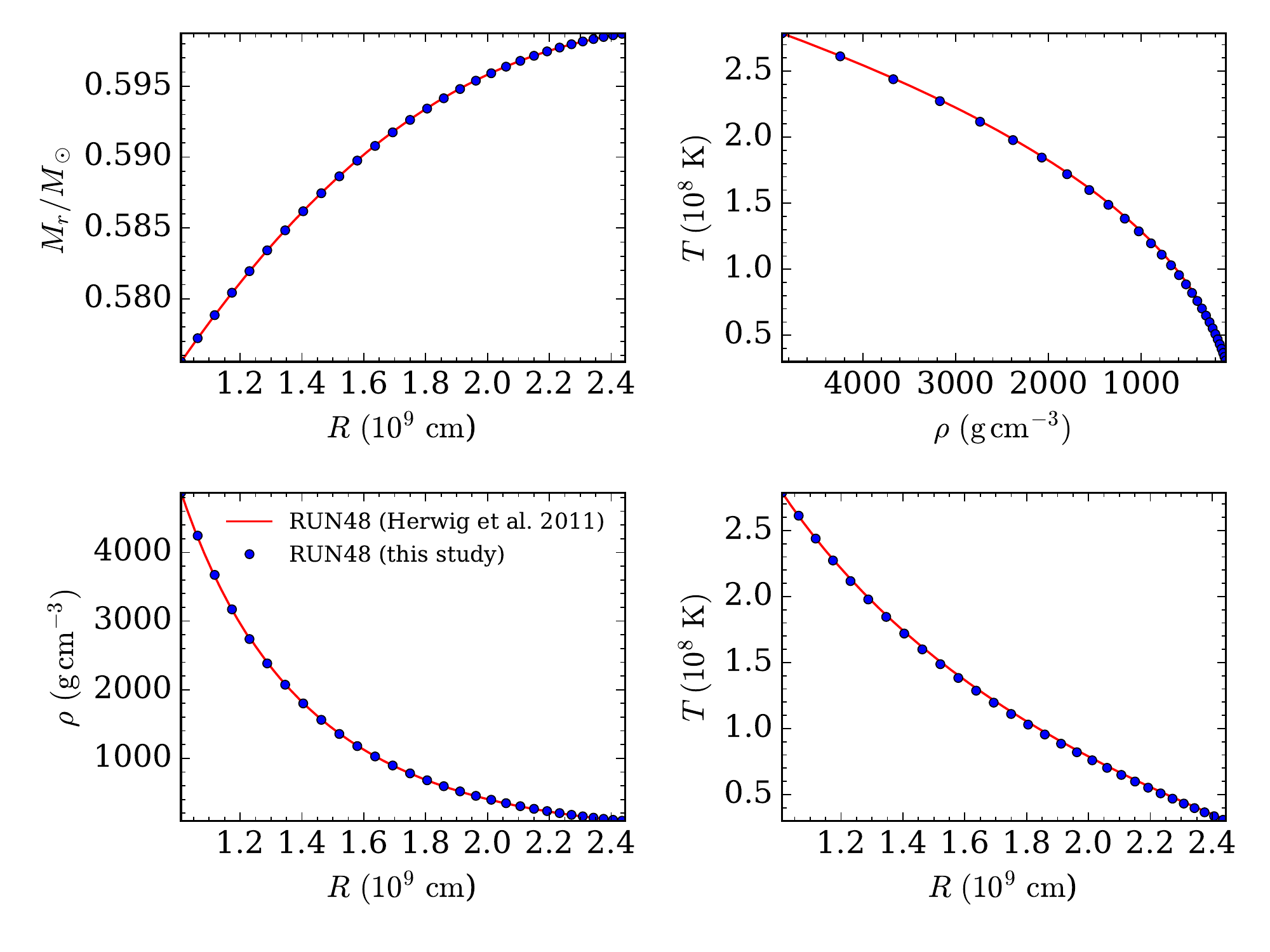}
  \caption{Comparison of the stellar structure parameter profiles in the He convective zones of RUN48 model from \citeasnoun{herwig:11}
           and RUN48 model implemented in this study (see text).
  }
  \label{fig:fig2}
\end{figure}

In the present study, RUN48 model is implemented with the following modifications. First, we take
$M_r$, $T$ and $\rho$ at the bottom of the He convective zone from the stellar evolution model ET14 
and obtain the $M_r$, $T$ and $\rho$ profiles as functions of $r$ in the He convective zone
by numerically integrating the equations of hydrostatic equilibrium and mass conservation
under the assumption of a constant entropy, as it should be in a deep convective zone, in the presence of an ideal gas and radiation. 
This is done with a view of our future stellar physics uncertainty and sensitivity study of
the models of Sakurai's object, because this approach allows to easily vary the properties of
the He-flash convection zone. Figure~\ref{fig:fig2} shows that the integrated profiles
match the original ones very well. Second, we use a higher resolution in the He shell with
132 mass zones instead of 81. These two modifications result in 13\% reduction of the i-process
simulation time after which our final abundances at the top of the He convective zone perfectly match their counterparts
from the original RUN48 model. Third, we average the final abundance distributions over
the entire He convective zone. Because the i-process nucleosynthesis takes place near the bottom of
the He convective zone, the produced (destroyed) heavy elements have profiles that are slightly
(because of convective mixing) increasing (decreasing) with the depth and time.
Therefore, the mass-averaging of the abundance profiles over the entire He zone allows us to
use a shorter (by 26\%) i-process simulation time before we reach the same final surface abundances as
in the original RUN48 model. Taking into account that we use a timestep of 74.3 sec versus 63 sec used
by \citeasnoun{herwig:11}, the total number of timesteps required to best match the chemical
composition of Sakurai's object is reduced from 1008 to 601, which significantly speeds up our Monte Carlo
simulations. Finally, we apply a decay time of 2 years to the final surface abundances of unstable isotopes,
which is much shorter than the 1 Gyr used in \citeasnoun{herwig:11} and matches the age of
Sakurai's object \cite{herwig:01}.

For the prepared (integrated and then interpolated for the specified mass zones) distributions of
$T$, $\rho$, $r$ and $D_\mathrm{conv}$ 
the mppnp code calculates the changes in the He-shell chemical composition due to nuclear reactions and accounts 
for mixing between the He and H zones. 
As in \citeasnoun{herwig:11}, the ingestion of H into the He-shell convection zone is taken into account in a parameterized way,
by adding $\Delta X = 10^{-4}$ of H mass fraction to the upper $4\times
10^{-4}\,M_\odot$ of the He convective zone at every 74.3 seconds, 
which corresponds to the H entrainment rate of $5.3\times 10^{-10} M_\odot\,\mathrm{s}^{-1}$
estimated from stellar evolution computations.

The initial abundances of the nuclear species affected by nuclear burning in the pp-chains, CNO cycle, and He-burning 
were taken from the stellar model calculations of \citeasnoun{herwig:99}. The dominant species are $X(^4\mbox{He})=0.35$,  $X(^{12}\mbox{C})=0.43$,
and $X(^{16}\mbox{O})=0.19$ \cite{herwig:99}, which reflects the primary production of $^{12}$C and $^{16}$O. For 
heavier elements we use the solar system abundance distribution of \citeasnoun{asplund:05} with isotopic ratios from \citeasnoun{lodders:03}
scaled to what would be expected for a metallicity (mass fraction of elements heavier than He) of $Z=0.01$ at the star's formation. 

We confirmed that the same initial composition can be obtained by taking the complete set of solar abundances of 
\citeasnoun{asplund:05} with isotopic ratios from \citeasnoun{lodders:03},  scaled to $Z=0.01$, and running a H-burning 
simulation  with the NuGrid one-zone code ppn with $T=55\times 10^6$ K and $\rho = 10^2\,\mathrm{g\,cm}^{-3}$, 
to complete H exhaustion,
followed by He burning at $T=150\times 10^6$ K and $\rho = 10^3\,\mathrm{g\,cm}^{-3}$, down to $X(^4\mbox{He})=0.35$.
This provides a cross-check of our initial abundances, and may be useful for future explorations of the impact of stellar
parameter variations. 

\begin{figure}
  \centering
  \includegraphics[width=10cm,viewport = 5 10 450 420]{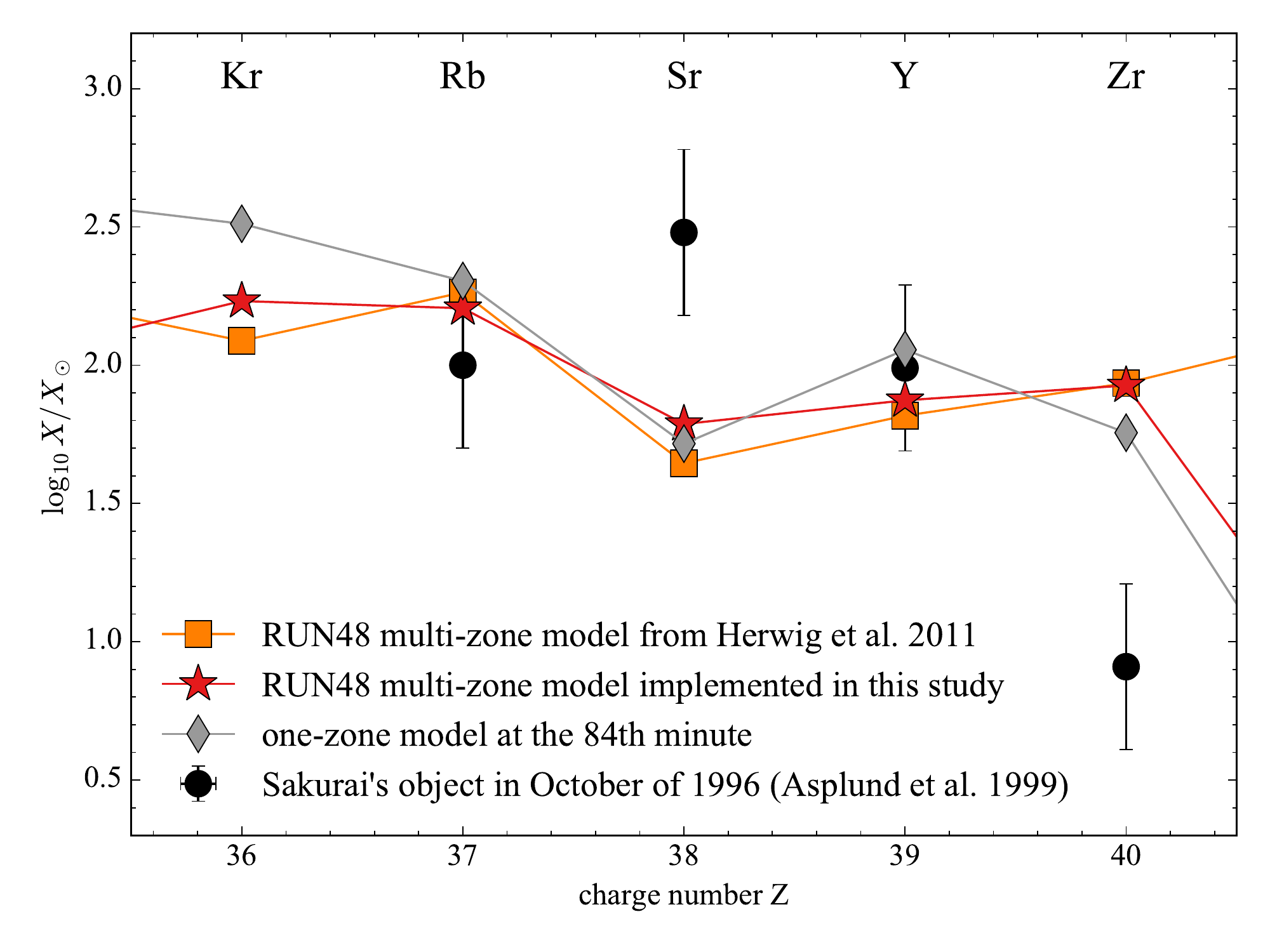}
  \caption{The observed (black circles with errorbars) and predicted (other symbols) solar-scaled abundances of 
  Rb, Sr, Y and Zr in Sakurai's
  object. The predicted abundances are from the multi-zone simulations terminated at the 1059th and 744th minutes
  for RUN48 model from \citeasnoun{herwig:11} and RUN48 model implemented in this study (see text), respectively,
  and from the 84th minute of nuclear burning in the one-zone simulations that best match the multi-zone results. 
  The small differences between the abundances from the two RUN48 models are entirely caused by the different decay
  times, 1 Gyr and 2 Yr, used in these models.
  }
  \label{fig:fig3}
\end{figure}

As in  \citeasnoun{herwig:11}, a key parameter for obtaining an abundance pattern that is similar to observations is the duration 
of i-process nucleosynthesis. In the stellar model, this duration is set by a split of the He-shell convection zone
that quenches the i-process. However, 
as discussed in  \citeasnoun{herwig:11}, the time of this split is highly uncertain given the limitations of 1D stellar models in accurately 
describing convective processes. Our goal here is to determine nuclear physics uncertainties for testing 
the hypothesis that the i-process is responsible for the observed abundances in Sakurai's object. We therefore 
determine the simulation time for our model (744 min) by requiring the best match of the He-shell mass-averaged abundances of Rb, Sr, 
Y, and Zr with observed abundances, after a 2 year decay time for long-lived radioactive isotopes 
that reflects the estimated lifetime of Sakurai's object \cite{herwig:01}. 
The resulting final abundances are shown in figure~\ref{fig:fig3}. 

\subsubsection{Nuclear reaction network and nuclear physics input}

\begin{figure}
  \centering
  \includegraphics[width=10cm,viewport = 5 10 450 420]{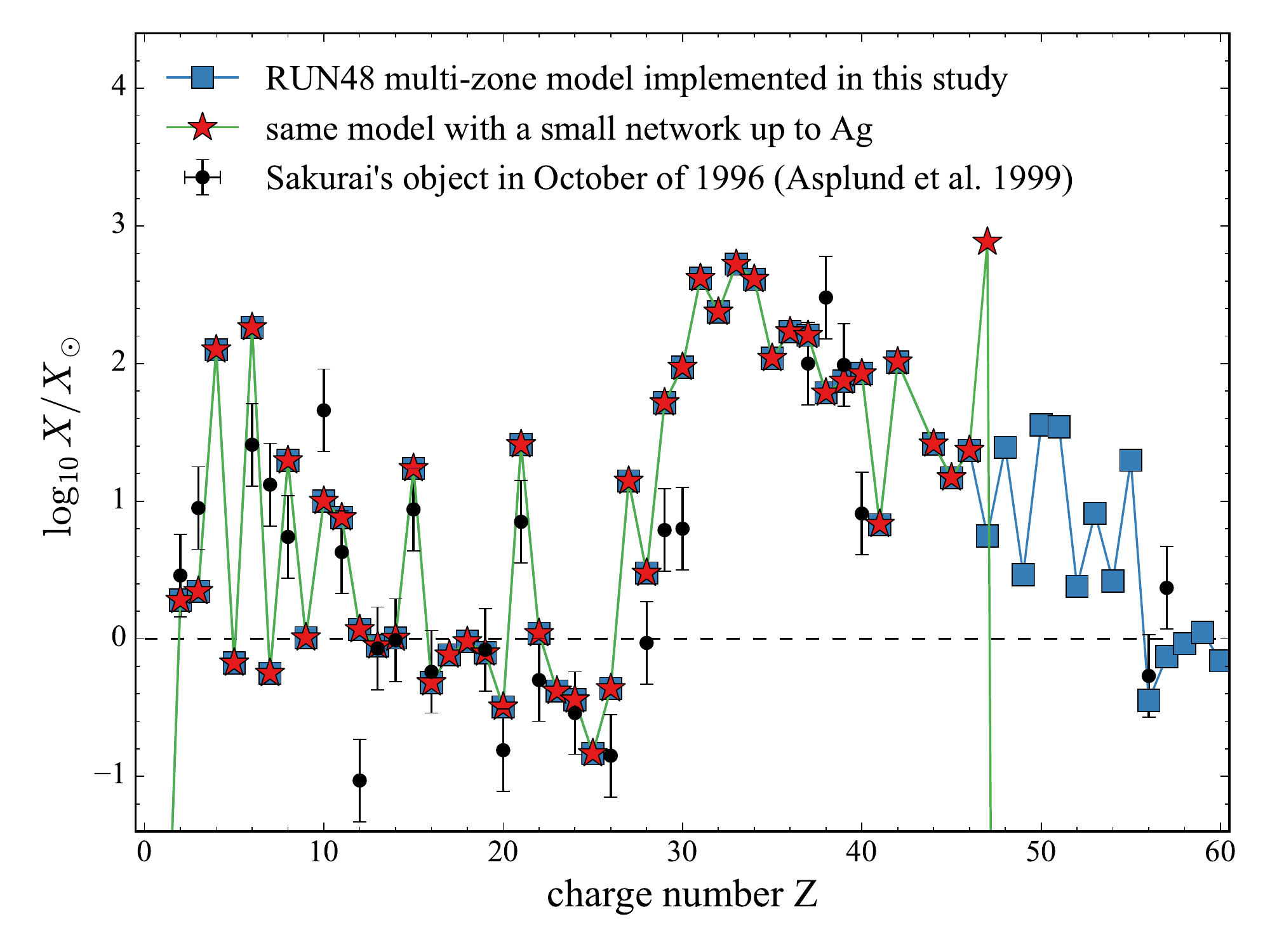}
  \caption{A comparison demonstrating that the multi-zone model used in this work reproduces reasonably well the global abundance
           distribution in Sakurai's object, in particular, the ratio of the abundances at the first and second s-process peaks.
           This comparison also shows that our using of a small network of 441 isotopes up to Ag does not affect 
           the abundances of the elements lighter than Ag, while allowing to significantly reduce the computational time.
  }
  \label{fig:fig4}
\end{figure}

The NuGrid nucleosynthesis network kernel that is used in both
the multi-zone and the single-zone (section \ref{sec:one-zone})
simulations extends the network dynamically based on the strength
of the nucleosynthesis fluxes, and may include up to $\sim
5{,}200$ isotopes and $\sim 67{,}000$ nuclear reactions \cite{herwig:08}. Two
parameters determine the dynamic network evolution. Species with
half lives below a minimum $\beta^-$-decay lifetime
of $10^{-3}$ seconds are not included. This sets the maximum width
of the network away from the valley of stability. Another parameter
sets the maximum mass of species to be included. Since in
this study we are interested in the first-peak elements, we save computing time
by not including species heavier than Ag. The maximum number of
species to be included in the network is then $441$. Limiting the
network in this way has no impact on the first-peak elements
(figure \ref{fig:fig4}).

Most reaction rates are taken from the JINA
REACLIB v1.1 library \cite{cyburt:10} with the following exceptions.
The NACRE recommended rates \cite{angulo:99} are used for all of the
CNO cycle reactions, except $^{13}$N(p,$\gamma)^{14}$O and
$^{14}$N(p,$\gamma)^{15}$O for which the JINA REACLIB rates are
used. The rates of the proton-induced reactions for isotopes with $A =
20$\,--\,40 are taken from \citeasnoun{iliadis:01}. For the
(n,$\gamma$) reactions with isotopes located along the valley of
stability, we mainly use the revision 0.3 of KADoNIS
data\footnote{\harvardurl{http://www.kadonis.org}} \cite{dillmann:06}
to linearly interpolate cross sections from the KADoNIS tables in energy,
more modern or different (n,$\gamma$) nuclear rates being used for
a sample of species\footnote{$^2$H \cite{nagai:06}, $^9$Be \cite{wallner:08}, 
$^{14}$C \cite{reifarth:08}, $^{17}$O \cite{wagoner:69}, $^{19}$F \cite{uberseder:07}, 
$^{21}$Ne \cite{heil:05}, $^{23}$Na \cite{heil:07}, $^{24,25,26}$Mg \cite{massimi:12}, 
$^{28,29,30}$Si \cite{guber:03}, $^{35,37}$Cl \cite{guber:02}, $^{40}$Ca \cite{dillmann:09}, 
$^{45}$Sc \cite{heil:09}, $^{54}$Fe \cite{coquard:06}, $^{58}$Fe, $^{59}$Co, 
$^{64}$Ni and $^{63,65}$Cu \cite{heil:08}, $^{60}$Fe \cite{uberseder:09}, 
$^{58}$Ni \cite{zugec:14}, $^{60}$Ni \cite{corvi:02}, $^{62}$Ni \cite{lederer:14}, 
$^{63}$Ni \cite{lederer:13}, $^{74,76}$Ge and $^{75}$As \cite{marganiec:09}, 
$^{78}$Se \cite{dillmann:06a}, $^{79,81}$Br and $^{85,87}$Rb \cite{heil:08a}, 
$^{80, 82, 83, 84, 86}$Kr \cite{mutti:05}, $^{90-96}$Zr \cite{lugaro:14}, $^{102}$Pd, 
$^{120}$Te, and $^{130,132}$Ba \cite{dillmann:10}, $^{116,120}$Sn \cite{koehler:01}, 
$^{138}$La (Dillmann 2006, priv. communication), $^{151,153}$Sn \cite{best:01}, 
$^{174}$Hf \cite{vockenhuber:07}, $^{168}$Yb, $^{180}$W, $^{184}$Os, $^{190}$Pt and 
$^{196}$Hg \cite{marganiec:10}, $^{184,186}$W \cite{marganiec:09a}, 
$^{185}$W \cite{mohr:04}, $^{186,187,188}$Os \cite{mosconi:07},  
$^{90,192}$Os (Marganiec 2008, priv. communication), $^{204}$Pb \cite{domingopardo:07}, 
$^{207}$Pb \cite{domingopardo:06b}, $^{209}$Bi \cite{domingopardo:06a}.}.
The compilations of \citeasnoun{fuller:85}, \citeasnoun{oda:94}, and
\citeasnoun{goriely:99} provide the NuGrid codes with weak reaction
rates.

\subsection{One-zone simulations}
\label{sec:one-zone}

We complement our multi-zone simulations with faster and simpler one-zone simulations 
and compare their results. For the one-zone simulations of the i process in Sakurai's object, we
use the same model as in
\citeasnoun{dardelet:14} employing the NuGrid one-zone nucleosynthesis code ppn. 
The one-zone simulation setup assumes constant
temperature and density, for which we adopt the typical He-shell
condition values of $T=2\times 10^8$ K and $\rho = 10^4$ g\,cm$^{-3}$. 

\begin{figure}
  \centering
  \includegraphics[width=10cm,viewport = 5 10 450 420]{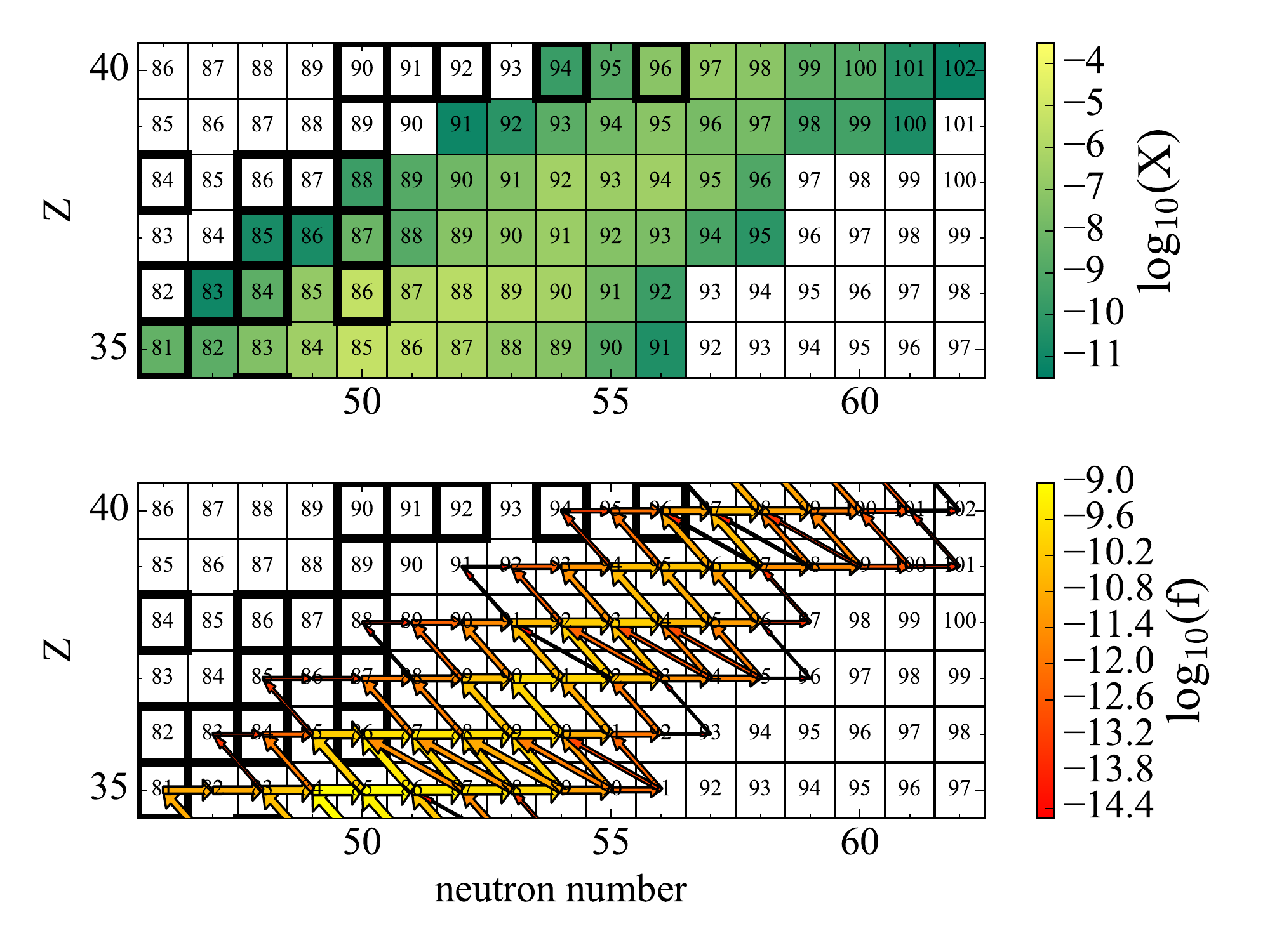}
  \caption{The mass fractions (upper panel) and reaction fluxes (lower panel) in the region of the chart of
           nuclides that surrounds the Rb, Sr, Y and Zr isotopes, shown for the 84th minute of our
           one-zone simulation of the i process in Sakurai's object. At this timestep of
           nuclear burning, the model has $\log_{10}\,N_\mathrm{n} = 15.65$ and $\tau = 1.01\,\mathrm{mbarn}^{-1}$.
           The nucleosynthesis flux, $f = (\delta Y_i/\delta t)_j$,
           shows the variation rate of the abundance $Y_i = X_i/A_i$ due to the reaction $j$. 
           The arrow width and color correspond to the flux strength. Heavy-lined boxes show the stable isotopes. 
             }
  \label{fig:fig5}
\end{figure}

Similar to the multi-zone case, the initial abundances for the
one-zone simulations are taken from the solar system abundance
distribution of \citeasnoun{asplund:05} with isotopic ratios from
\citeasnoun{lodders:03} scaled to
$Z=0.01$.  In the one-zone model, we adopt $X(^1\mbox{H})=0.1$, which
is supposed to mimic the H ingestion into the He convective zone, and
$X(^{12}\mbox{C})=0.5$ at the expense of $^{16}$O, assuming that the new $^{16}$O mass fraction
is equal to the sum of the old mass fractions of $^1$H, $^{12}$C and $^{16}$O minus the sum of the new mass fractions of $^1$H and $^{12}$C. 
The adopted H abundance is much higher than that expected in any zone of more realistic 1D or
3D multi-zone simulations.  However, this initial
condition in the one-zone model reproduces the neutron densities of the i
process \cite{dardelet:14}.

All of our one-zone nucleosynthesis simulations of the i process 
in Sakurai's object were run with the same lists of isotopes and nuclear reactions and
used the same reaction rates as their corresponding multi-zone simulations.
We also used the same 2 year decay time after the end of the one-zone simulations.

The grey diamonds in figure~\ref{fig:fig3} represent the abundances from the one-zone simulations at the 84th minute
that best match the final abundances of the multi-zone simulations. Figure~\ref{fig:fig5} shows 
the mass fractions and reaction fluxes in the region of the chart of nuclides that includes
the stable (inside the heavy-line boxes) and unstable isotopes of Br, Kr, Rb, Sr, Y and Zr.

\subsection{Nuclear reaction rate variations}

The dependence of the abundance (mass fraction $X_k$) predictions  on reaction rate variations $\Delta r_i$ for $n$ rates $r_i$ is given by the Taylor series
\begin{eqnarray}
X_k(r_1+\Delta r_1,\ldots,r_{n}+\Delta r_{n}) = & X_k(r_1,\ldots,r_{n}) +
\sum_{i=1}^{n}\frac{\partial X_k}{\partial r_i}\Delta r_i + \nonumber \\
& \frac{1}{2}\sum_{i=1}^{n}\sum_{j=1}^{n}\frac{\partial^2 X_k}{\partial r_i \partial r_j}\Delta r_i\Delta r_j + \ldots.
\label{eq:taylor}
\end{eqnarray}
We implement the rate variations using multiplication factors $f_i$, in which case
$\Delta r_i = r_i(f_i-1)$, where $f_i \geq 0$. 

In order to determine the nuclear physics uncertainties, the stellar model calculations are repeated for different 
sets of nuclear reaction rate variations. In this work we focus on systematic uncertainties in theoretical predictions of reaction rates (see section \ref{results}).
We therefore represent the estimated uncertainty of a reaction rate with a maximum variation factor $v_i^\mathrm{max} > 1$. The reaction rate $r_i$ can then 
take any value between $r_i/v_i^\mathrm{max}$ and $r_i v_i^\mathrm{max}$ with equal probability density. 
In our Monte-Carlo (MC) simulation setup we partly follow \citeasnoun{stoesz:03} and \citeasnoun{rauscher:16}. 
First, we randomly select for each reaction $i$ a variation factor $v_i^\mathrm{rand}$ from its uniform distribution in
the interval $1\leq v_i^\mathrm{rand} \leq v_i^\mathrm{max}$. The rate multiplication factor used is then 
$f_i = (p/v_i^\mathrm{rand}) + (1-p)v_i^\mathrm{rand}$, where $p$ is randomly chosen to be either 0 or 1 with equal probability.

This selection method is used to generate $m=10{,}000$ sets of multiplication factors $\{f_{i,j}\}_{j=1}^{m}$
and for each set a multi-zone post-processing 
calculation is carried out. As a result one obtains for each element 10{,}000 abundance values $X_{k,j}$. The distribution of these values
can be fitted with a normal distribution to obtain mean values $\mu$ and standard deviations $\sigma$. $\sigma$ can be interpreted 
as the nuclear physics related uncertainty of the abundance value. We have verified that 10{,}000 is indeed a sufficiently large number of MC sets
by comparing results to a calculation based on 5{,}000 sets. The resulting values of $\mu$ and $\sigma$ change by less than 1\%.
In order to identify the most important sources of nuclear physics uncertainties, we follow \citeasnoun{rauscher:16} 
and search for correlations between the varied multiplication factors $f_i$ and predicted elemental abundances
$ X_k$ by calculating the Pearson product-moment correlation coefficient 
\begin{eqnarray}
r_\mathrm{P}(f_i,\,X_k) = \frac{\sum_{j=1}^{m}(f_{i,j}-\overline{f_i})(X_{k,j}-\overline{X_k})}
{\sqrt{\sum_{j=1}^{m}(f_{i,j}-\overline{f_i})^2}\sqrt{\sum_{j=1}^{m}(X_{k,j}-\overline{X_k})^2}},
\label{eq:rP}
\end{eqnarray}
where $\overline{f_i} = (\sum_{j=1}^{m}f_{i,j})/m$ and similarly for $\overline{X_k}$.

While the MC approach is suitable to identify the dominant nuclear uncertainties, there is the possibility that smaller, 
but still significant, sources of uncertainty are missed, especially when a large number of reactions contributes 
with comparable strength. We therefore complemented our MC study 
with a series of simulations D1, in which only one reaction rate is changed per run. 
For each reaction rate, two runs are carried out with the maximum multiplication 
factor $f_i = v_i^\mathrm{max}$ (``rates up'') and the minimum multiplication 
factor $f_i = (1/v_i^\mathrm{max})$ (``rates down''). This approach determines 
to first order the partial derivatives $\frac{\partial X_k}{\partial r_i}$ in equation \ref{eq:taylor}. 
In order to get an understanding of the significance of correlations between reaction rate changes, 
we also carried out a complete set of calculations D2 with variations of all possible reaction rate pairs. 
These calculations determine all $\frac{\partial^2 X_k}{\partial r_i \partial r_j}$ partial derivatives in equation \ref{eq:taylor}.

\section{Nuclear physics input uncertainties} 
\label{physics}

For a given neutron density evolution (which is affected by its own nuclear uncertainties from the neutron producing reactions, 
and nuclear uncertainties in the prior stellar evolution) i-process nucleosynthesis calculations rely on neutron capture rates 
and $\beta^-$ decay rates within the i-process band, which extends from two to eight
neutrons away from the valley of stability (figure~\ref{fig:fig5}). In this mass range the terrestrial $\beta^-$-decay rates used in this work are all 
experimentally well known, though corrections for the stellar environment 
may be significant in some cases.
However, none of the neutron capture rates on unstable nuclei 
are experimentally constrained. We therefore focus in this study on the role of uncertainties in theoretical neutron capture rate predictions. 

Neutron capture rates are predicted using the Hauser-Feshbach model of statistical decay of a
compound nucleus. This theoretical model describes
the decay of ``highly'' excited nuclei with large numbers of levels per
MeV, provided that a reasonable description of nuclear statistical properties (for
example level density and gamma ray strength function) and nuclear potentials are available. Two sources of
uncertainty need to be distinguished (1) the uncertainty due to the inherent limitations of the statistical model 
that describes an expected average behavior of nuclei (intrinsic uncertainty) and (2) the uncertainty due to 
the inability of models for nuclear potentials, level densities, and gamma-ray strength functions to describe the 
average behavior of nuclei away from stability where experimental data on these quantities are not available (extrapolation uncertainty). 

The intrinsic model error can be constrained by comparison of rate predictions with experimental reaction rate data in 
mass regions where experimental data constrain global descriptions of level density, gamma ray strength functions and nuclear potentials. 
Currently reliable measurements of astrophysical neutron capture rates are only possible for stable or very long-lived 
isotopes. A comprehensive comparison of neutron capture rate predictions from statistical models with experimental data was 
carried out by \citeasnoun{beard:14} and demonstrated that the intrinsic uncertainty of the statistical model approach 
is of the order of a factor of 2. 

The extrapolation uncertainty for predictions of neutron capture rates away from stability in the i-process region 
is more difficult to estimate. We follow here the approach of \citeasnoun{liddick:16} who took 
advantage of the fact that for each of the statistical model ingredients (nuclear potentials, level densities, 
and gamma-ray strength functions) there is a range of theoretical models that describe the 
data near stability reasonably well. The divergence of the predictions of these various models away from stability can then 
be taken as a lower limit of the model uncertainty, which can then be propagated to the nuclear reaction rates. 
\citeasnoun{liddick:16} found that for r-process temperatures of 1.5~GK within a few mass units away 
from stability rate uncertainties in the Mn-Ga range quickly exceed factors of 10 or even 100. 
We apply here the same approach to the nuclide region and lower temperatures ($2\times 10^8$ K) of relevance for the i-process in Sakurai's object. 

We use the statistical model code TALYS\footnote{\harvardurl{http://talys.eu}} \cite{TALYS:07}
and take advantage of various options for potentials, nuclear 
level densities (NLD) and $\gamma$ ray strength functions ($\gamma$SF). The choice of input models was restricted 
to the models listed in table~\ref{table:local_potentials}.  
The selection criteria are discussed in more detail in \citeasnoun{liddick:16}. Comparison calculations with the Koning-Delaroche \cite{Kon03a} and 
Jekeune-Lejenne-Mahaux 
\cite{Jeu77a} optical potentials yielded no significant differences. 
We therefore used the Koning-Delaroche potential for all calculations. 

\begin{figure}
  \centering
  \includegraphics[width=13cm,viewport = 5 10 550 480]{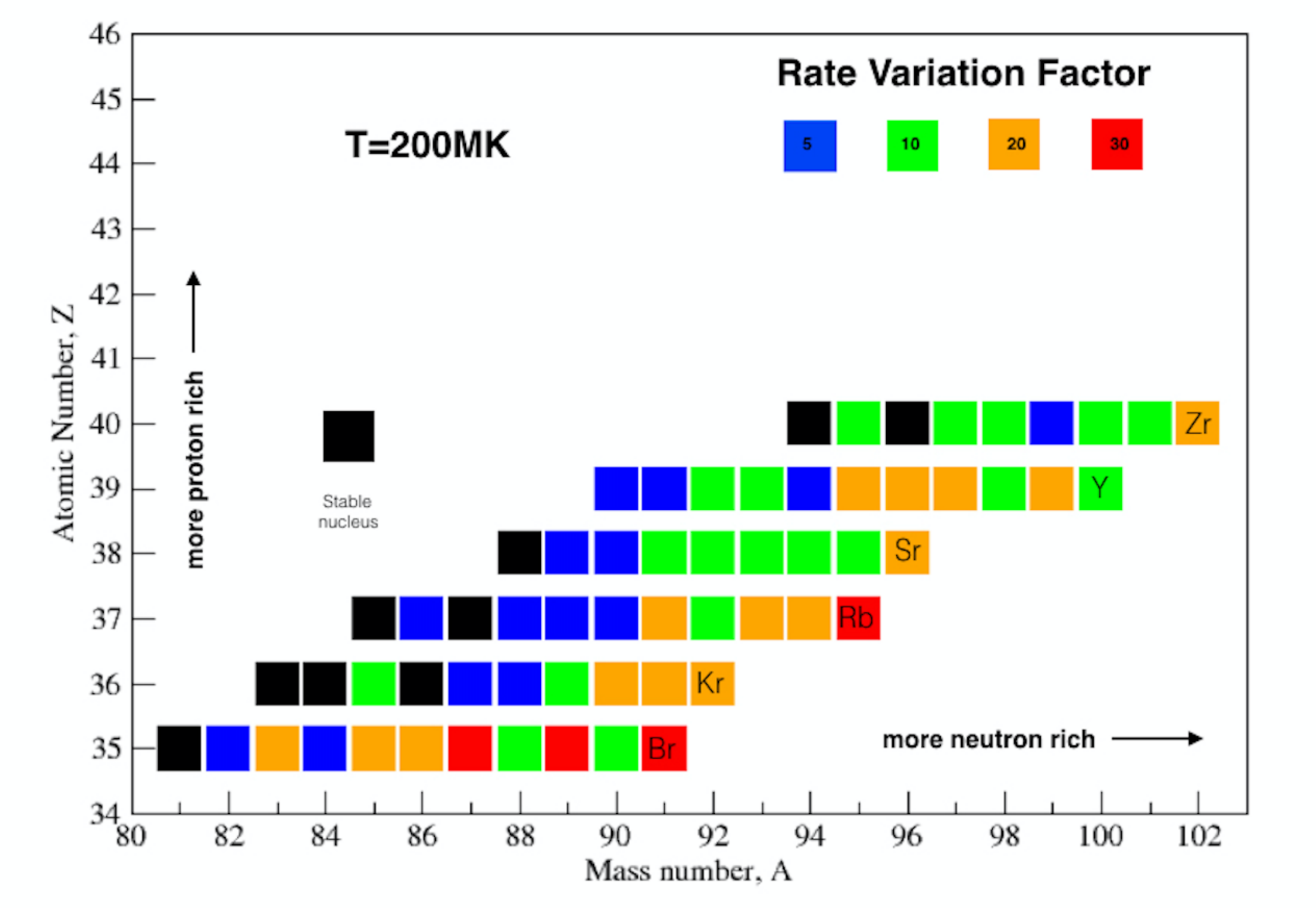}
  \caption{The maximum variation factor $v_i^\mathrm{max}$ used for each of the 52 unstable isotopes considered in this
  study based on variations in statistical model predictions of neutron capture rates using various combinations of input models. 
  } 
  \label{fig:fig6}
\end{figure}

Calculations were carried out with all combinations of the 5 NLD models and 4 $\gamma$SF models in table~\ref{table:local_potentials}. 
The ratio of the largest and smallest neutron capture rate is indicated
in figure~\ref{fig:fig6} and increases from less than a factor of 5 to more than a factor of 20 for increasingly neutron rich nuclei. 
We bin the ratio into ranges and use the values indicated in figure~\ref{fig:fig6} as an estimate of the extrapolation uncertainty in the reaction rate 
predictions. As it is generally much larger than the intrinsic uncertainty of about a factor of 2, we use this ratio as 
the maximum variation factor $v_i^{max}$ in our sensitivity study, $v_i^{max}=30$ being used for $v_i^{max}>20$. 

\begin{table}
\caption{\label{table:local_potentials}List of models used to describe the nuclear level density and 
the $\gamma$ strength in the Hauser-Feshbach calculations. Calculations were performed with all 
20 possible combinations of these models.}
\label{tab:HFmodels}
\begin{indented}     
\item[]\begin{tabular}{@{}c}
\br
Nuclear Level Density (NLD) models used in this work \\
\mr    
Constant Temperature matched to the Fermi Gas (CT+BSFG) \cite{Dil73a} \\
Back-shifted Fermi Gas (BSFG) \cite{Dil73a,Gil65a} \\
Generalized Super fluid (GSM) \cite{Ign79a,Ign93a} \\
Hartree Fock using Skyrme force (HFS) \cite{Gor01a} \\
Hartree-Fock-Bogoliubov + combinatorial (HFBS-C) \cite{Gor08a} \\
\br
\end{tabular}

\medskip
\item[]\begin{tabular}{@{}c}
\br
$\gamma$ ray Strength Function ($\gamma$SF) models used in this work\\
\mr 
Kopecky-Uhl generalized Lorentzian (KU) \cite{Kop90a}\\
Hartree-Fock BCS $+$ QRPA (HF-BCS+QRPA) \cite{Gor02a}\\
Hartree-Fock-Bogolyubov $+$ QRPA (HFB+QRPA) \cite{Gor04a}\\
Modified Lorentzian (Gor-ML)\cite{Gor98a}\\
\br
\end{tabular}
\end{indented}
\end{table}

The subset of neutron capture reactions varied in this study (see figure~\ref{fig:fig6}) was identified using the 
single-zone model. We include 52 neutron capture rates that 
exhibit strong reaction fluxes $\log_{10} f > -14.4$ at the peak of 
i-process processing just prior to its quenching by the split of the convection zone (figure~\ref{fig:fig5}). 
$^{90}$Y was included for completeness. 

\section{Results}\label{results}

In this section we present the simulation results for varying 
the radiative n-capture rates of the 52 unstable isotopes shown in figure~\ref{fig:fig6} 
both randomly and systematically. We analyze the impact of these reaction rate 
variations on the predicted surface abundances of
Rb, Sr, Y and Zr in models of Sakurai's object. 

\subsection{Multi-zone Monte-Carlo simulations}
\label{sec:mc}

\begin{figure}
  \centering
  \includegraphics[width=10cm,viewport = 5 10 450 420]{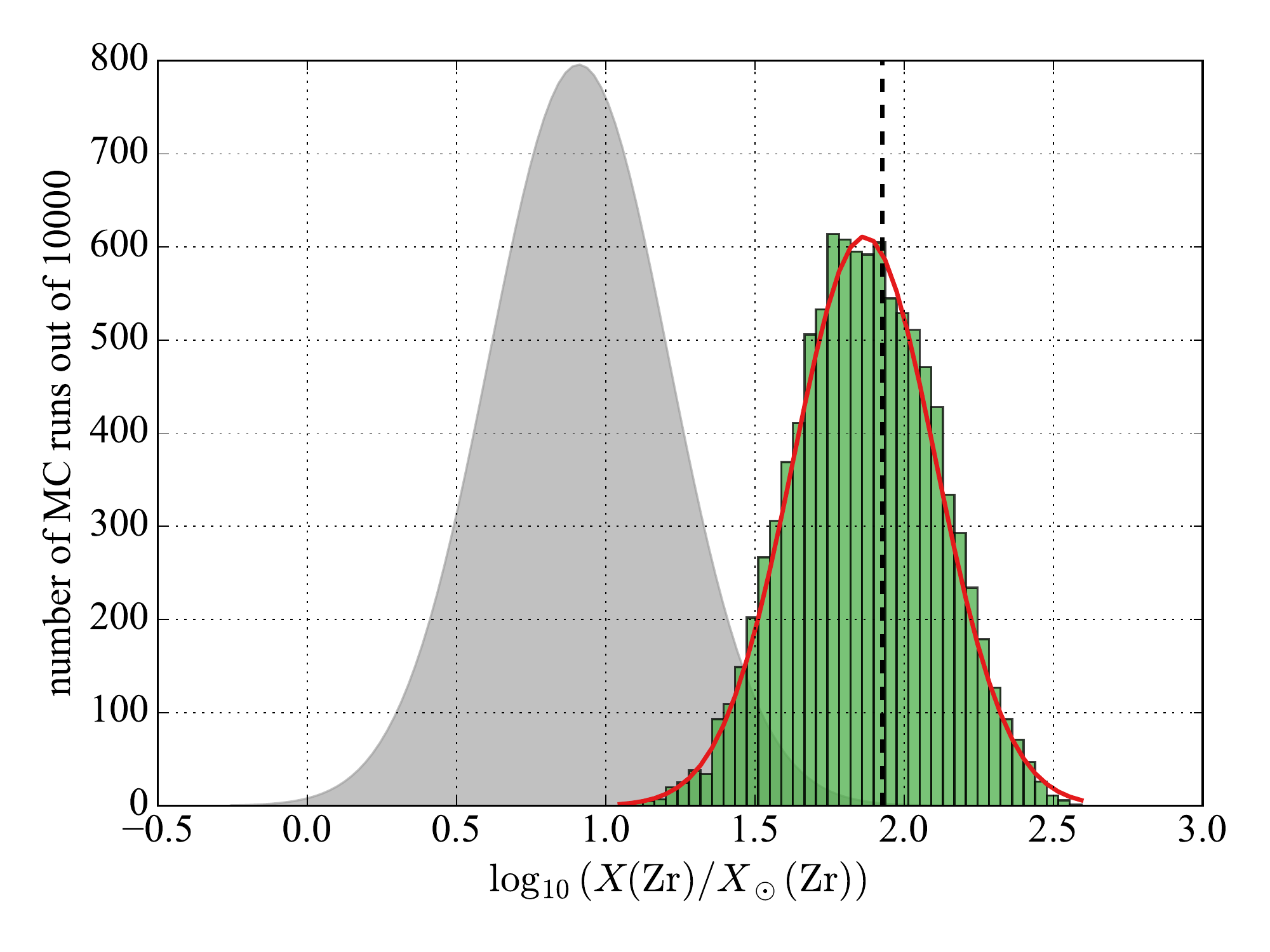}
  \caption{A distribution of the predicted surface mass fraction of Zr from our MC simulations (the green histogram). 
    The vertical dashed line is the benchmark model prediction for the reaction rate multiplication factors $f_i = 1$. 
    By fitting a normal distribution (the red curve), we estimate the mean $\mu$ and standard deviation $\sigma$ of the results. 
    The benchmark model prediction is shifted relative to the mean of the fitted normal distribution
    because $\log_{10}\langle X\rangle = \mu + 0.5\sigma^2$ for the lognormal distribution of $X$.
    For comparison, the grey-shaded area shows the observed Zr abundance 
    represented by a normal distribution with the mean and standard deviation from \citeasnoun{asplund:99}.
  } 
  \label{fig:fig7}
\end{figure}

\begin{figure}
  \centering
  \includegraphics[width=10cm,viewport = 5 10 450 420]{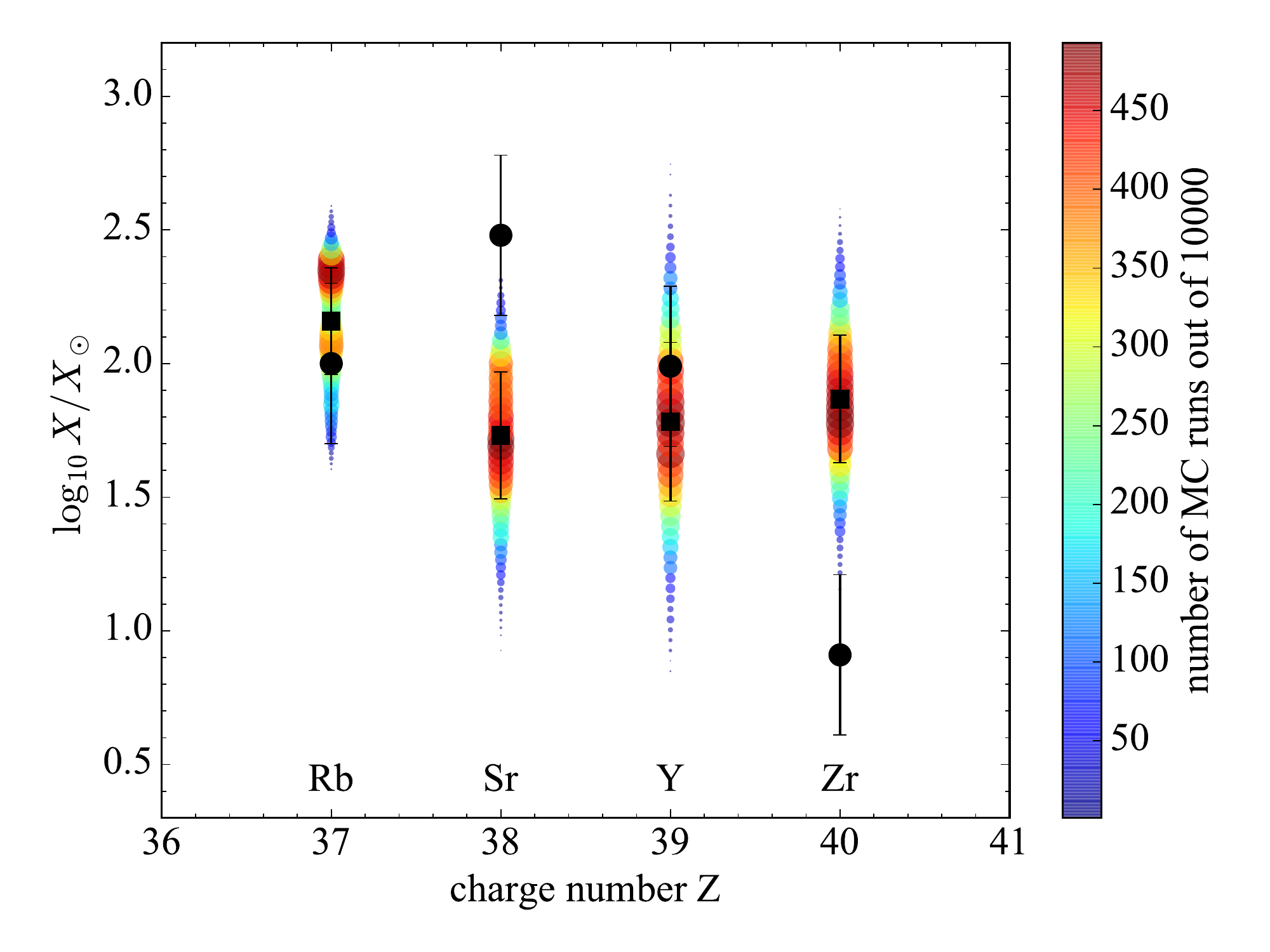}
  \caption{The circle color- and size-coded distributions of the predicted surface mass fractions of 
    Rb, Sr, Y and Zr in Sakurai's object from our MC simulations. The squares with error bars show 
    the mean values and standard deviations estimated by fitting with
    normal distributions, as illustrated in figure~\ref{fig:fig7}. These distributions and error bars indicate the nuclear physics 
    uncertainties in the model predictions. Observational data are shown as black circles with error bars \cite{asplund:99}.
  } 
  \label{fig:fig8}
\end{figure}

\begin{figure}
  \centering
  \includegraphics[width=10cm,viewport = 5 10 450 420]{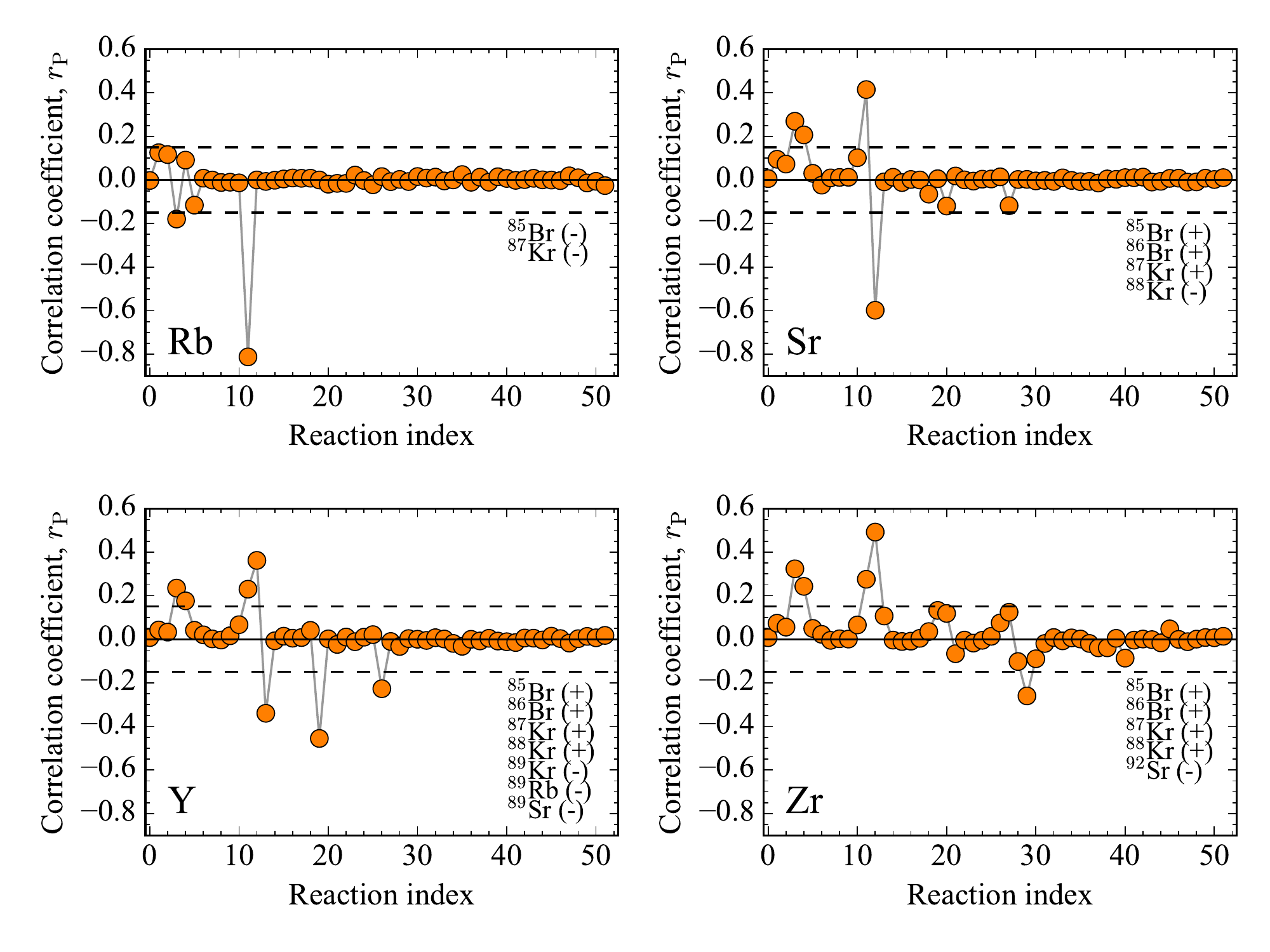}
  \caption{The Pearson correlation coefficients (equation \ref{eq:rP})
           for (anti-) correlations between all the 52 (n,$\gamma$) reaction rates
           selected for this study and the predicted elemental abundances of Rb, Sr, Y and Zr. The isotopes with $r_\mathrm{P}\leq -0.15$
           and $r_\mathrm{P}\geq 0.15$ (below $(-)$ and above $(+)$ the dashed lines) are indicated for each of the four elements.
  }
  \label{fig:fig9}
\end{figure}

\begin{figure}
  \centering
  \includegraphics[width=10cm,viewport = 5 10 450 420]{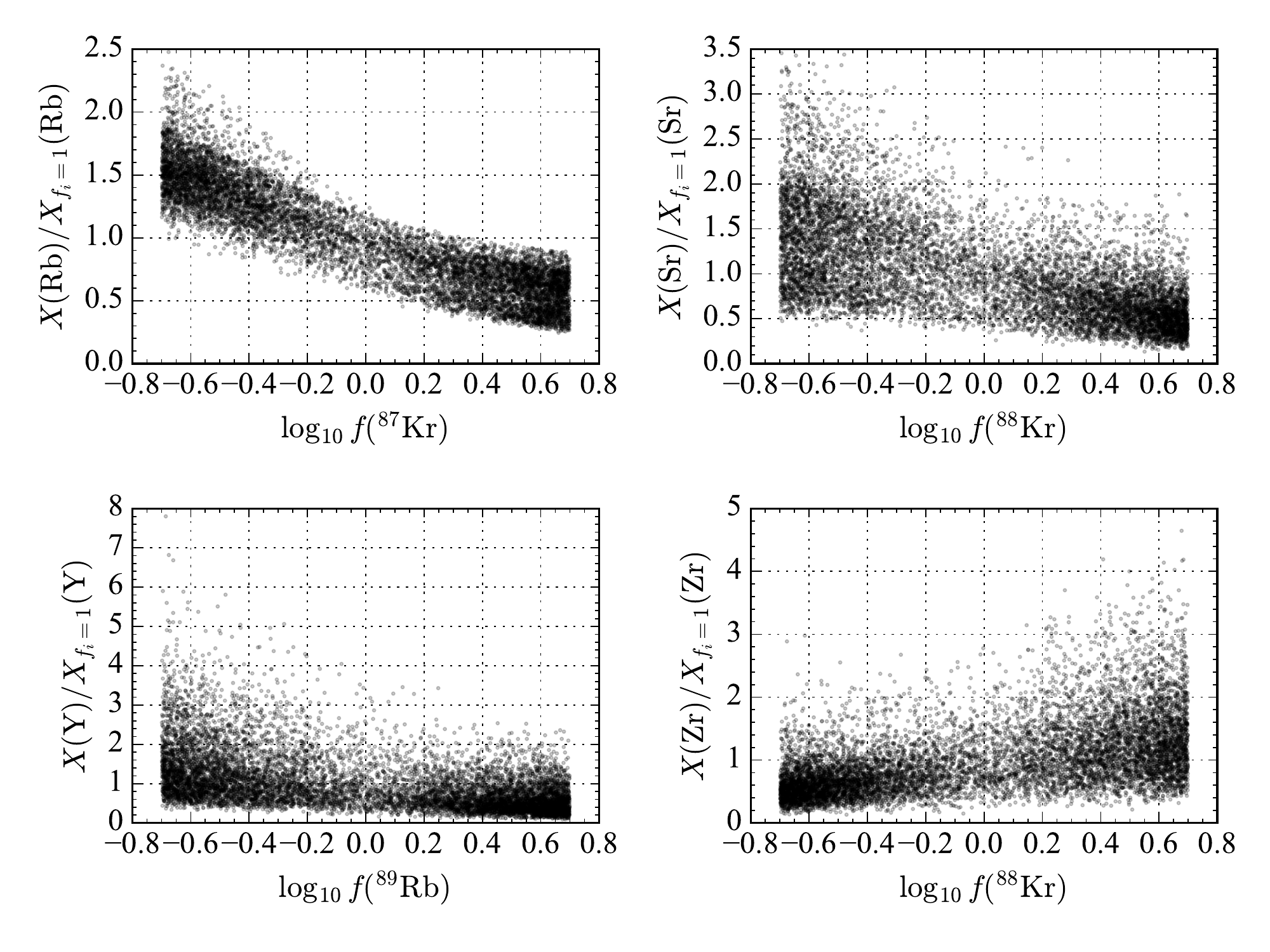}
  \caption{The ratios of the final abundances of Rb, Sr, Y and Zr relative to the benchmark case of $f_i=1$
    as functions of the multiplication factors for the reactions with
    the maximum $|r_\mathrm{P}|$ from table \ref{tab:tabcorr}.
  }
  \label{fig:fig10}
\end{figure}

Most of the abundance distributions from the 10{,}000 Monte Carlo variations of the 52 (n,$\gamma$) reaction rates follow closely a normal distribution. 
As an example, figure~\ref{fig:fig7} shows the distribution of the predicted surface abundance of Zr 
(logarithmic and with respect to its solar value). We therefore fitted the distributions with a normal distribution
to obtain standard deviation and mean. Figure~\ref{fig:fig8} shows the Monte Carlo predicted abundance distributions, 
together with mean and standard deviations from a fit with a normal distribution, in comparison with 
observations from Sakurai's object. It is evident from this figure that nuclear uncertainties are significant and 
overall comparable to observational errors.

\begin{table}
\caption{The Pearson product-moment correlation coefficient (equation \ref{eq:rP}) for the reactions
that have (anti-) correlations between the rate multiplication factor and the predicted elemental
abundances of Rb, Sr, Y and Zr and their ratios with $|r_\mathrm{P}| > 0.15$ in our MC multi-zone simulations of the i process in Sakurai's object.}
\label{tab:tabcorr}
\begin{indented}
\item[]\begin{tabular}{@{}cccccccc}
\br
Reaction & Rb & Sr & Y & Zr & Rb/Sr & Sr/Y & Y/Zr \\
\mr
$^{85}$Br(n,$\gamma$) & $-0.18$ & $+0.27$ & $+0.23$ & $+0.32$ & $-0.22$ & & \\
$^{86}$Br(n,$\gamma$) &  & $+0.21$ & $+0.18$ & $+0.24$ & & & \\
$^{87}$Kr(n,$\gamma$) & $-0.81$ & $+0.41$ & $+0.23$ & $+0.28$ & $-0.60$ & & \\
$^{88}$Kr(n,$\gamma$) & & $-0.60$ & $+0.36$ & $+0.49$ & $+0.42$ & $-0.59$ & \\
$^{89}$Kr(n,$\gamma$) & & & $-0.34$ & & & $+0.27$ & $-0.37$ \\
$^{89}$Rb(n,$\gamma$) & & & $-0.46$ & & & $+0.42$ & $-0.54$ \\
$^{89}$Sr(n,$\gamma$) & & & $-0.23$ & & & $+0.21$ & $-0.30$ \\
$^{92}$Sr(n,$\gamma$) & & & & $-0.26$ & & & $+0.21$ \\
\br
\end{tabular}
\end{indented}
\end{table}

(Anti-) correlations between the neutron capture reaction rate
variations and the final abundances of Rb, Sr, Y and Zr indicate particularly important reaction rate uncertainties (figure \ref{fig:fig9}). 
The most significant (anti-) correlations are shown in figure \ref{fig:fig10}, and are for
absolute magnitudes of $|r_\mathrm{P}|\geq 0.15$ listed in table~\ref{tab:tabcorr} and in figure \ref{fig:fig9}. 
Rb is very strongly affected by a single reaction rate 
uncertainty from $^{87}$Kr(n,$\gamma$). This explains the deviation of its resulting abundance distribution from a normal distribution (figure \ref{fig:fig8}). 
Sr is affected by mainly 2 uncertainties --- $^{87}$Kr(n,$\gamma$) and $^{88}$Kr(n,$\gamma$). Y  and Zr are affected by a larger number of reactions with comparable 
$|r_\mathrm{P}|$. Table~\ref{tab:tabcorr} also shows that some reactions have a strong impact on elemental abundance ratios. 
For example, $^{87}$Kr(n,$\gamma$) affects strongly Rb/Sr, $^{88}$Kr(n,$\gamma$) affects strongly Sr/Y, 
and $^{89}$Rb(n,$\gamma$) affects strongly Y/Zr. 

Some of these results can be anticipated if we compare the mass fractions of isobars
from the one-zone simulation presented in upper panel of figure~\ref{fig:fig5}. From such a comparison, one may come to the following
conclusions. The final abundance of Zr could have been most affected by the production of $^{90}$Kr if its decay
to $^{90}$Zr had not been shielded by the long-lived $^{90}$Sr whose half life time $t_{1/2}\approx 28.8$ yr considerably
exceeds our used 2 yr decay time. Therefore, the Zr abundance in Sakurai's object is actually determined by the synthesis of $^{92}$Sr. 
The results of our statistical analysis in table~\ref{tab:tabcorr} confirm this conclusion. The presence of $^{85}$Br,
$^{86}$Br, $^{87}$Kr and $^{88}$Kr with the positive correlation coefficients in the Zr, Y and Sr columns of 
table~\ref{tab:tabcorr} rather indicates that the n-capture cross sections of these isotopes determine the total strength of the i process in our model.
$^{89}$Y is the only stable isotope of Y. It is mainly produced from the decay of the short-lived $^{89}$Kr.
The only unshielded stable isotope of Sr is $^{88}$Sr. It is predominantly produced from $^{88}$Kr that has the maximum
mass fraction among the isobars with $A=88$. Rb has two unshielded stable isotopes, $^{85}$Rb and $^{87}$Rb, but the production of
$^{85}$Rb via the $\beta^-$-decay of the abundant $^{85}$Br is hindered because it passes through the relatively long-lived ($t_{1/2}\approx 10.8$ yr)
isotope $^{85}$Kr. Therefore, the final Rb abundance is determined by the synthesis of $^{87}$Rb from $^{87}$Kr, which
agrees with the results presented in table~\ref{tab:tabcorr}.

\subsubsection{Multi-zone variations of individual rates (D1 and D2)}
\label{sec:d1}

We also carried out 104 multi-zone model calculations where each of the 52 reaction rates was varied up and down by 
the highest ($f_i = v_i^\mathrm{max}$) and lowest ($f_i = 1/v_i^\mathrm{max}$) variation factor, 
respectively (calculation set D1). Based on this simplified approach we find the sensitivities listed in tables~\ref{tab:tabymaxD1}
and \ref{tab:tab2}. Almost all the important reaction rate 
uncertainties identified with the Monte Carlo study are also identified as significant sensitivities 
in the single rate variations when we select the reactions for which the condition $\log_{10}|X_{f_i\neq 1}/X_{f_i=1}| \ge 0.075$ is true
for at least one of the final abundances of Rb, Sr, Y and Zr. 
There is only one additional reaction of comparable sensitivity in the single rate variations --- $^{85}$Kr(n,$\gamma$) ---
that does not show a correlation with $|r_\mathrm{P}|>0.15$ in the Monte Carlo study.
This indicates that single rate variations are as sensitive in identifying sources of uncertainty
as the correlation analysis of the Monte Carlo results. 

\begin{table}
\caption{The list of reactions (the first column) for which the absolute magnitude of
at least one of the logarithmic ratios of the abundances of Rb, Sr, Y and Zr predicted with
the rate multiplication factor $f_i$ varied up and down relative to the case of $f_i=1$ (shown in the other columns)
has exceeded $0.075$ in our multi-zone D1 runs.}
\label{tab:tabymaxD1}
\begin{indented}
\item[]\begin{tabular}{@{}c|c|c|c|c}
\br
Reaction & Rb (up/down) & Sr (up/down) & Y (up/down) & Zr (up/down) \\
\mr
$^{85}$Br(n,$\gamma$) & $-0.102$/$0.028$ & $0.068$/$-0.029$ & $0.07$/$-0.03$ & $0.071$/$-0.03$ \\
$^{86}$Br(n,$\gamma$) & $0.034$/$-0.006$ & $0.068$/$-0.014$ & $0.073$/$-0.015$ & $0.077$/$-0.016$ \\
$^{85}$Kr(n,$\gamma$) & $-0.005$/$0.007$ & $0.016$/$-0.093$ & $0.016$/$-0.092$ & $0.016$/$-0.094$ \\
$^{87}$Kr(n,$\gamma$) & $-0.225$/$0.231$ & $0.104$/$-0.28$ & $0.085$/$-0.21$ & $0.07$/$-0.157$ \\
$^{88}$Kr(n,$\gamma$) & $0.0$/$0.0$ & $-0.305$/$0.185$ & $0.151$/$-0.269$ & $0.145$/$-0.239$ \\
$^{89}$Kr(n,$\gamma$) & $0.0$/$0.0$ & $0.0$/$0.0$ & $-0.276$/$0.066$ & $0.045$/$-0.017$ \\
$^{89}$Rb(n,$\gamma$) & $0.0$/$0.0$ & $0.003$/$0.005$ & $-0.226$/$0.241$ & $0.038$/$-0.089$ \\
$^{89}$Sr(n,$\gamma$) & $0.0$/$0.0$ & $0.006$/$-0.007$ & $-0.088$/$0.121$ & $0.013$/$-0.027$ \\
$^{92}$Sr(n,$\gamma$) & $0.0$/$0.0$ & $0.0$/$0.0$ & $0.0$/$0.0$ & $-0.089$/$0.117$ \\
\br
\end{tabular}
\end{indented}
\end{table}

We also varied all possible pair combinations of the 52 neutron capture reactions to explore correlations 
between uncertainties (calculation  set D2). This required 5304 additional multi-zone calculations. 
The results of the D2 simulations are presented in the second column of table~\ref{tab:tab2}.
As expected, the maximum changes in final abundance are larger when two reactions are changed simultaneously. 
However, the list of important nuclear physics uncertainties is the same, again indicating that correlations 
among rate changes are negligible, and that for the case of the i-process a single rate variation approach is 
appropriate for identifying the critical nuclear reactions. 

\begin{table}
\caption{\label{table_results}Individual isotopes (D1) and their pairs (D2) whose maximum (n,$\gamma$) reaction rate variations have the largest impact
on the predicted elemental abundances and their ratios (the absolute magnitudes are given in parenthesis) 
from multi-zone simulations of the i process in Sakurai's object.}
\label{tab:tab2}
\begin{indented}
\item[]\begin{tabular}{@{}cc}
\br
D1 & D2 \\
\mr
$^{87}$Kr ($\Delta X (\mathrm{Rb}) = 0.23$\,dex) & $^{85}$Br and $^{87}$Kr ($\Delta X (\mathrm{Rb}) = 0.49$\,dex) \\
$^{88}$Kr ($\Delta X (\mathrm{Sr}) = 0.31$\,dex) & $^{87}$Kr and $^{88}$Kr ($\Delta X (\mathrm{Sr}) = 0.53$\,dex) \\
$^{89}$Kr ($\Delta X (\mathrm{Y}) = 0.28$\,dex) & $^{88}$Kr and $^{89}$Rb ($\Delta X (\mathrm{Y}) = 0.53$\,dex) \\
$^{88}$Kr ($\Delta X (\mathrm{Zr}) = 0.24$\,dex) & $^{88}$Kr and $^{89}$Rb ($\Delta X (\mathrm{Zr}) = 0.34$\,dex) \\
\mr
$^{87}$Kr ($\Delta [\mathrm{Rb}/\mathrm{Sr}] = 0.51$) & $^{87}$Kr and $^{88}$Kr ($\Delta [\mathrm{Rb}/\mathrm{Sr}] = 0.76$) \\
$^{88}$Kr ($\Delta [\mathrm{Sr}/\mathrm{Y}] = 0.46$) & $^{88}$Kr and $^{89}$Rb ($\Delta [\mathrm{Y}/\mathrm{Sr}] = 0.72$) \\
$^{89}$Rb ($\Delta [\mathrm{Y}/\mathrm{Zr}] = 0.33$) & $^{89}$Kr and $^{89}$Rb ($\Delta [\mathrm{Zr}/\mathrm{Y}] = 0.58$) \\
\br
\end{tabular}
\end{indented}
\end{table}

\begin{table}
\caption{\label{table_sz_results}Same as in Table~\ref{table_results}, but from one-zone simulations.}
\label{tab:tab3}
\begin{indented}
\item[]\begin{tabular}{@{}cc}
\br
D1 & D2 \\
\mr
$^{85}$Br ($\Delta X (\mathrm{Rb}) = 0.18$\,dex) & $^{85}$Br and $^{87}$Br ($\Delta X (\mathrm{Rb}) = 0.62$\,dex) \\
$^{85}$Br ($\Delta X (\mathrm{Sr}) = 0.36$\,dex) & $^{85}$Br and $^{88}$Kr ($\Delta X (\mathrm{Sr}) = 0.63$\,dex) \\
$^{89}$Kr ($\Delta X (\mathrm{Y}) = 0.65$\,dex) & $^{85}$Br and $^{89}$Kr ($\Delta X (\mathrm{Y}) = 1.15$\,dex) \\
$^{85}$Br ($\Delta X (\mathrm{Zr}) = 0.44$\,dex) & $^{85}$Br and $^{88}$Kr ($\Delta X (\mathrm{Zr}) = 0.78$\,dex) \\
\mr
$^{85}$Br ($\Delta [\mathrm{Rb}/\mathrm{Sr}] = 0.52$) & $^{85}$Br and $^{87}$Kr ($\Delta [\mathrm{Rb}/\mathrm{Sr}] = 0.85$) \\
$^{89}$Kr ($\Delta [\mathrm{Sr}/\mathrm{Y}] = 0.70$) & $^{88}$Kr and $^{89}$Kr ($\Delta [\mathrm{Y}/\mathrm{Sr}] = 1.02$) \\
$^{89}$Kr ($\Delta [\mathrm{Y}/\mathrm{Zr}] = 0.78$) & $^{89}$Br and $^{89}$Kr ($\Delta [\mathrm{Zr}/\mathrm{Y}] = 0.90$) \\
\br
\end{tabular}
\end{indented}
\end{table}

\subsection{One-zone simulations}

The D1 and D2 series of simulations have been repeated using the one-zone model of the i process in Sakurai's object.
The results are presented in table~\ref{tab:tab3}. The one-zone simulations
identify neutron capture rates on $^{85}$Br, $^{87}$Br, $^{87}$Kr, $^{88}$Kr, and $^{89}$Kr as the dominant uncertainties
affecting the final abundances. Although this list partially overlaps with the one obtained 
from multi-zone simulations, it over-emphasizes the importance of $^{87}$Br(n,$\gamma$) and overlooks $^{86}$Br(n,$\gamma$),
$^{89}$Rb(n,$\gamma$), $^{89}$Sr(n,$\gamma$), and $^{92}$Sr(n,$\gamma$). We conclude that single zone models are not adequate for 
i-process sensitivity studies of convective-reactive processes, such as the i process in Sakurai's object. 

\section{Summary and Conclusions}
\label{s.conclusions}

We studied the impact of nuclear uncertainties on the nucleosynthesis predictions of a multi-zone i-process model 
for the post-AGB star Sakurai's object to determine whether this model can explain the large observed enhancements 
in Rb, Sr, Y, and Zr. We analyzed the impact of the uncertainties in the theoretical predictions of 52 (n,$\gamma$) 
reaction rates on neutron rich unstable isotopes of Br-Zr with neutron numbers around $N=46-61$. 
These reactions cannot be measured directly in experiments as both reactants, the neutron rich isotope and the neutron, 
are unstable. We estimated the theoretical uncertainties by 
varying input parameters in the TALYS nuclear statistical model code and find, 
in agreement with previous work \cite{liddick:16} that uncertainties increase dramatically with distance 
from the valley of stability due to the lack of experimental constraints on key ingredients, 
such as level structure and $\gamma$-strength distribution. These reaction rate uncertainties can reach more 
than a factor of 10 across the band of the i-process. 

We propagated these errors through the stellar model using a Monte Carlo approach and the NuGrid 1D multi-zone 
post-processing framework mppnp that enables sufficiently fast modeling with large nuclear reaction networks, 
while taking fully into account convective mixing processes. We find that the nuclear uncertainties 
in the predicted abundances can reach 0.3 dex and are thus comparable to current observational uncertainties. 
With both, observational and nuclear uncertainties quantified, we are now in a position to compare abundance predictions 
with observations and test the appropriateness of the underlying astrophysical model (see figure~\ref{fig:fig3}). 
Model predictions agree well with observations for Rb and Y. Abundance predictions are low for Sr, and high for Zr. 
However, they do agree with observations at the 2$\sigma$ level. To determine whether these differences are significant 
and point to issues with the astrophysical model assumptions  would require a significant reduction of 
uncertainties in either observations, nuclear physics, or both. 

There are significant opportunities to reduce the nuclear physics uncertainties. Experimental techniques 
have been developed to indirectly constrain neutron capture rates on unstable nuclei. Neutron transfer reactions 
\cite{cizewski:07,bardayan:16} and other surrogate techniques \cite{escher:12} populate compound nuclei and 
study their properties. The $\beta$-Oslo technique uses $\beta^-$-decay to constrain level densities and 
$\gamma$-strength functions and has specifically been developed to reduce uncertainties in statistical model 
predictions of neutron capture rates on unstable nuclei \cite{liddick:16}. Furthermore, significant increases 
in intensities of  radioactive beams needed for these measurements
are expected with new rare isotope beam facilities such as FRIB in the US or FAIR in Germany coming online in the near future.

To guide future experimental and theoretical efforts in reducing nuclear physics uncertainties in i-process models 
we identified the most critical reactions that are the dominant sources of uncertainties. The reactions identified using 
the Pearson product-moment correlation coefficient between rate and abundance changes in the Monte Carlo calculations, 
and reactions identified by single variation of individual reaction rates mostly agree. This demonstrates that single rate 
variations are an appropriate approach to identify critical reaction rates. The important reaction rate uncertainties 
for i-process production of Rb, Sr, Y, and Zr are neutron captures on $^{85}$Br, $^{86}$Br, $^{85}$Kr, 
$^{87}$Kr, $^{88}$Kr, $^{89}$Kr, $^{89}$Rb, $^{89}$Sr, and $^{92}$Sr. Neutron captures on $^{85}$Kr were 
only identified in the single rate variations, indicating that their uncertainties are less dominant than the others. 
Single rate variations using a simplified one-zone i-process led to significantly different results. 
This shows that one-zone approximations are not reliable for identifying critical reaction rates 
in convective-reactive regimes such as the post-AGB star i-process. 

The 1D stellar model used here has significant astrophysical uncertainties. For example, the split of the convection zone 
that quenches the i-process, and therefore actually determines its duration, had to be delayed compared 
to what has been predicted by the mixing length theory 
in the 1D stellar model in order to obtain a strong i-process production of Rb, Sr, Y, and Zr \cite{herwig:11}. 
The goal here was to investigate whether comparison with observations reveals any deficiencies of 
the underlying astrophysical model that would guide future model improvements to address astrophysical model uncertainties. 
We found that with current nuclear uncertainties model predictions and observations are consistent, 
though there are some possible hints of discrepancies. Future improvements in nuclear uncertainties are needed 
to provide strong guidance for model improvements. In addition, it will be important to investigate 
the sensitivity to astrophysical parameters such as H-ingestion rate and duration, $r$, $T$ and $\rho$ 
at the bottom of the convective zone, $Z$, $D_\mathrm{conv}$, or mass resolution. 
Should future work reveal strong significant discrepancies with observations, such sensitivity studies can 
then point to specific issues in the model. 

Our results were obtained using an i-process model for the post-AGB star Sakurai's object. 
The model does reproduce the observed abundances within nuclear and observational errors. 
We therefore expect our nuclear sensitivity results to be applicable to any post-AGB star i-process model 
that reproduces the observed abundances of Rb, Sr, Y, and Zr. Our results should also be directly applicable 
to near-solar metallicity rapidly accreting white dwarfs, where models have indicated that an i-process occurs 
under similar physical conditions \cite{denissenkov:17}. The computational method and analysis tools that we have developed and used 
in conjunction with the NuGrid post-processing 
nucleosynthesis codes can be used for uncertainty and sensitivity studies of many other convective-reactive nucleosynthesis sites. 

\ack This material is based upon work supported by the National
Science Foundation under Grant No. PHY-1430152 (JINA Center for the
Evolution of the Elements).  M.~P. acknowledges the support from the
``Lend\"{u}let-2014'' Programme of the Hungarian Academy of Sciences
and from SNF (Switzerland).  F.~H. receives NSERC funding through a
Discovery grant.
S.~J. is a fellow of the Alexander von Humboldt Foundation and acknowledges support from the Klaus Tschira Stiftung.
The computations for this work were made on the computer Mammouth Parall\`{e}le 2 from the University of Sherbrooke, 
managed by Calcul Qu\'{e}bec and Compute Canada. The operation of this computer is funded by the Canada Foundation 
for Innovation (CFI), the minist\`{e}re de l'\'{E}conomie, de la science et de l'innovation du Qu\'{e}bec (MESI) and 
the Fonds de recherche du Qu\'{e}bec – Nature et technologies (FRQ-NT). H.S. and A.S. acknowledge support from the National 
Science Foundation under Grant No. PHY-1102511. 

\section*{References}

\bibliography{paper}
\bibliographystyle{jphysicsB}
\end{document}